\documentclass[aps,floats,prl,showpacs,twocolumn ]{revtex4}

\usepackage{amsfonts,amsmath} \usepackage{bm} \usepackage{dcolumn}
\usepackage{epsfig} \usepackage{latexsym}

\begin{document}

\title{Universal high-frequency transport in perfect photonic crystals}

\author{Chushun Tian$^{1,2}$ and Luwei Zhou$^3$}

\affiliation{$^1$ Institut f{\"u}r Theoretische Physik,
Universit{\"a}t zu K{\"o}ln, K{\"o}ln, 50937, Germany\\
$^2$ Center for Nonlinear Studies, Hongkong Baptist University,
Kowloon Tong, Hongkong\\
$^3$ National key laboratory of surface physics and Department of
physics, Fudan University, Shanghai, 200433, P. R. China}

\date{\today}
\begin{abstract}
 {\rm The light scattering in the periodic dielectric
 cylinder array is studied. We analytically calculate the diffusive-ballistic transport
 crossover and find the weak localization superimposing on it.
 Possible experimental observations are analyzed.
 }
\end{abstract}


\maketitle

{\it Introduction}---Experimental studies \cite{Fishman07,Maret06}
are offering new boost to interest in transport of optical systems.
A traditional subject is the light scattering in fully
\cite{Maret06} or partially \cite{Fishman07,Vos00} disordered media.
Irrespective of the absence/presence of the periodic background
disorders turn out to be essential in both kinds of materials.
There, the Anderson localization of light originates at the multiple
disorder scattering. The latter leads to the enhanced constructive
interference between two counterpropagating optical paths and, thus,
suppresses the light diffusion.

For perfect periodic dielectric materials (namely photonic crystals)
the low-frequency transport is protocoled by sample-specific band
structures, the analysis of which is nowadays a well-established
industry. By contrast, the high-frequency transport in perfect
photonic crystals is a largely unexplored subject which might find
practical applications. There, localization and periodicity may
interplay with each other. Theoretically, no localization signatures
have been found for periodic quantum mapping \cite{Dorfman04}, which
might suggest that localization is incompatible with perfect
periodicity. Contrary to this, for realistic quantum particle motion
in some perfect crystals, the weak localization correction to the
classical diffusion coefficient is analytically shown \cite{Tian05}.
Experimentally, to detect localization signals in periodic
structures, if any, turns out to be extremely difficult. Indeed, in
usual systems (e.g. electrons) ubiquitous particle interactions
complicate analysis of wave interference effects such as (weak)
localization.
Although this problem is circumvented in optical systems,
traditional transmission/reflection measurements do not allow
directly probing the dynamical (bulk) diffusion coefficient.

\begin{figure}
 \begin{center}
 \leavevmode \epsfxsize=6cm \epsfbox{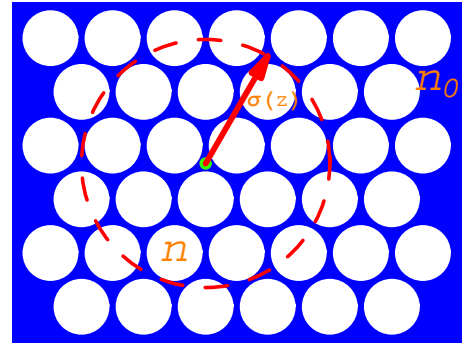}
 \end{center}
 \caption{A narrow input probe beam (green spot)--perpendicular to the plane--is launched into
 the dielectric cylinder array and expands. The ``time'' axis points inwards.}
  \label{section}
\end{figure}

The recent experiment \cite{Fishman07} offers the very opportunity
to study the diffusive-ballistic transport crossover. The
experimental scope consists of two crucial steps: (i) to fabricate
the dielectric material which is uniform in the longitudinal
direction (but might be arbitrary in the transverse plane) and, (ii)
to launch a probe laser beam, parallel to the longitudinal
direction, into the material (Fig.~\ref{section}). Then, the
dynamical process of the dispersion in the transverse plane is
recorded by the intensity profile at different propagating
distances, which allows the direct measurement of the dynamical
diffusion coefficient. It is the purpose of this letter to
analytically study the crossover in perfect photonic crystals and to
show that weak localization superimposes on this crossover.
Furthermore, we examine the possibility of confirming these
predictions within this experimental scope.

{\it Qualitative results based on band structure}---To be specific
we will consider the following photonic crystals. Dielectric
cylinders of a radius $r_0$\,, uniform in the ``longitudinal''
($z$-) direction, are periodically embedded into the host material
of a refractive index $n_0$ forming a triangular lattice in the
``transverse'' ($x$-$y$) plane with a lattice constant $a$
(Fig.~\ref{section}). For this lattice let us choose the coordinate
system as $(x,y)={\bf l} + {\bf r}$\,, where ${\bf l} \in
\mathfrak{B}$ (the Bravis lattice) is chosen to be the cylinder
center, and ${\bf r}$ specifies the position inside the unit cell. A
probe beam of frequency $\Omega_0$\,, parallel to the $z$-axis, is
launched into the host dielectric material, then expands as
propagates along the $z$-direction. The transverse wave number
$k_\perp$ is inversely proportional to the initial width of the
beam, which is narrow enough such that $k_\perp^{-1}\ll \min
(a-2r_0, r_0)$\,.

Then, the propagation of the scalar field (for simplicity) denoted
as $E(x,y,z)$ inside the photonic crystal is described by the
Helmholtz equation (The light velocity in the vacuum is set to be
unity.): $[\nabla^2 + \Omega_0^2 n^2(x,y)]E(x,y,z)=0$
with $n(x,y)$ the refractive index profile. Introduce the
slow-varying envelope $\psi(x,y;z)$ such that $E(x,y,z) \equiv
\psi(x,y;z)\, e^{i\Omega_0 n_0z}$ and insert it into the Helmholtz
equation. In the paraxial limit $|\partial_z^2 \psi|\ll \Omega_0 n_0
|\partial_z \psi| $ the second order derivative may be ignored
\cite{Fishman07}, consequently we arrive at the Schr{\"o}dinger
equation:
\begin{equation}
\left\{i\, \partial_z + \frac{1}{2\Omega_0 n_0}\, (\partial_x^2 +
\partial_y^2) - V(x,y) \right\}\psi(x,y;z) = 0 \,,
\label{wave}
\end{equation}
where $V(x,y)\equiv -\frac{\Omega_0}{2n_0} [n^2(x,y)-n_0^2]$\,. This
shows that the beam propagates in the manner fully analogous to the
two-dimensional quantum particle motion in the ``potential''
$V(x,y)$\,, and $z$ plays the role of the usual time.

In periodic potentials exists the band structure $\epsilon_n ({\bf
k})$\,, where $n$ is the band index and ${\bf k}$ is the
quasimomentum in the first Brioullin zone
($\mathbb{B}.\mathbb{Z}.$). As to be detailed below (for a large
class of potentials \cite{potential}) the ray optics inside a unit
cell is ergodic characterized by the positive Lyapunov exponent
$\lambda$\,. Following the well-known Bohigas-Giannoni-Schmit (BGS)
conjecture, then, given ${\bf k}$ (away from symmetry points or
lines) the high-frequency bands display universal level fluctuations
\cite{Mucciolo94}.

If a single band is excited the motion of (Bloch) photons is
ballistic. However, because the beam initially is narrow enough it
excites a number of high-frequency bands centered at $\epsilon=
k_\perp^2/(2\Omega_0 n_0)$ with an amount of $\sim \lambda/\Delta
\gg 1$\,, where $\Delta$ is the mean level spacing. At early times
the rays are successively bent by dielectric cylinders which,
equivalently, may be viewed as that Bloch photons experience
considerable inelastic scattering. Consequently, the beam dispersion
(expansion), defined as $\sigma^2 (z)\equiv \int\!\!\!\!\int dxdy
(x^2+y^2) |\psi(x,y;z)|^2$\,, is dominated by the interband
transition factor: $e^{i \{\epsilon_n ({\bf k})-\epsilon_{n'} ({\bf
k})\}z}$\,. Then, the large number of excited bands allows us to
average this factor with respect to the universal level correlator,
which is $\sim \{\epsilon_n ({\bf k})-\epsilon_{n'} ({\bf
k})\}^{-2}$ \cite{Efetov97,Tian08b}. Consequently, the normal
diffusion, i.e., $\sigma^2 (z)\sim z$ is recovered for early times
$\lambda^{-1}\lesssim z \ll \Delta^{-1}$\,. At later times $z\gtrsim
\Delta^{-1}$ individual bands are resolved and the interband
transition becomes negligible. The transport, in turn, becomes
ballistic, i.e., $\sigma^2 (z)\sim \Delta z^2$ \cite{Tian08b}.

For intermediate times $\lambda^{-1} \ll z \lesssim \Delta^{-1}$
photons move in the regime of size $\sim \sqrt{D_{cl}/\Delta}$\,,
where the periodic structure is not resolved and the dielectric
cylinders thereby resemble random scatterers. As usual while two
optical paths diffusively propagate they find a significant
probability to form a common loop and counter-propagate along it.
Such constructive interference suppresses the normal diffusion
coefficient $D_{cl}$ and the weak localization correction results,
which is order of $(\nu D_{cl})^{-1}$ with $\nu$ the photon density
of states. Accordingly, the linear dispersion is suppressed. That
is,
\begin{eqnarray}
\sigma^2(z) \sim \bigg\{ \begin{array}{cc}
  z \,, \qquad \qquad \qquad & z \gtrsim \lambda^{-1}\,, \\
  z \left(1 - \frac{\ln \lambda z}{\nu D_{cl}} \right)\,, & \lambda^{-1} \ll z \lesssim \Delta^{-1}\,, \\
  \Delta z^2\,,\qquad \qquad \qquad & z\gtrsim \Delta^{-1}\,.
\end{array}
\label{widthresult}
\end{eqnarray}

We now turn to the detailed analysis of wave dynamics described by
Eq.~(\ref{wave}) and derive the rigorous results. Below a finite
system composed of ${\cal N}$ unit cells will be considered with the
periodic boundary condition implemented. In the final results the
limit: ${\cal N}\rightarrow +\infty$ is set.

{\it Field-theoretic formalism}---First we present the general
formalism used throughout this work. Consider the propagation on the
lattice from ${\bf l}$ to ${\bf l}'$\,, described by the so-called
two-point correlator: ${\cal Y}({\bf l}-{\bf l}';\omega) \equiv
{\cal N}\int \!\!\!\!\int \! d{\bf r}d{\bf r}'
G^R_{\epsilon+\frac{\omega}{2}}\left({\bf r}+{\bf l}, {\bf r}'+{\bf
l}'\right)G^A_{\epsilon-\frac{\omega}{2}}\left({\bf r}'+{\bf
l}',{\bf r}+{\bf l}\right)$\,, noticing that ${\bf r}$\,, ${\bf r}'$
are the coordinates in the unit cell. Here $G^{R/A}$ is the
retarded/advanced Green function of Eq.~(\ref{wave}) in the
frequency domain. The translational symmetry implies that ${\cal Y}$
depends merely on ${\bf l}-{\bf l}'$\,.

Following Ref.~\cite{Tian08a} the Green functions are written down
explicitly in the Bloch basis $|n{\bf k}\rangle$\,. In doing so the
wave dynamics is reduced into the one in a two-torus. For the latter
we invoke the Wigner transform and pass to the phase space ${\mathbb
T}$\,. The coordinate is denoted as $X \equiv ({\bf r},{\bf p})$
(after appropriate rescaling) and satisfies $\int_{\mathbb T}
dX=1$\,. Then, it is a canonical procedure
\cite{Efetov97,Tian08a,Andreev96} to express the spatial Fourier
transform of ${\cal Y}({\bf l}-{\bf l}';\omega)$\,, denoted as
${\cal Y}({\bf q},\omega)$\,, in terms of the functional integral
over the supermatrix field $Q=T\Lambda T^{-1}$\,, where both $Q$ and
$T$ depend on $X$\,. Moreover, $Q(X)={\rm K}{\rm C}^{\rm T}Q^{\rm
T}({\bar X}){\rm C}{\rm K}$ and $T^\dagger(X)={\rm C}T^{\rm T}({\bar
X}){\rm C}^{\rm T}$ with ${\bar X}\equiv ({\bf r},-{\bf p})$\,.
After tedious calculations we find (Throughout this work we are
interested in the low-lying modes, i.e., ${\bf q}\rightarrow 0$\,.)
\begin{eqnarray}
{\cal Y}({\bf q},\omega) = \frac{(\pi\nu)^2}{64 {\cal N}}
\tilde{\sum_{{\bf k}_\pm}}\int\!\!\!\!\int _{{\mathbb T}} \!\!
dXdX'e^{i{\bf q}\cdot ({\bf r}-{\bf r}')}
\!\!\int \!\! DQ e^{-F[Q,{\bf a}]} \nonumber\\
\times {\rm str}
\left[k(1+\Lambda)(1-\tau_3)Q(X)k(1-\Lambda)(1-\tau_3)Q(X')\right]\,,
\label{propagator1}
\end{eqnarray}
where ${\tilde \Sigma}_{{\bf k}_\pm}\equiv \Sigma_{{\bf k}_\pm \in
{{\mathbb{B}.\mathbb{Z}.}}}\delta_{{\bf k_+}-{\bf k}_-,{\bf q}}$\,,
${\rm str}$ is the supertrace, and the action is
\begin{eqnarray}
F[Q,{\bf a}]=\frac{\pi\nu}{2}\int_{\mathbb T} dX \, {\rm str}
\left\{T\Lambda {\hat {\cal L}}_{\rm a}T^{-1} + \frac{i\omega^+}{2}
\Lambda Q\right\}\,. \label{freeenergy}
\end{eqnarray}
Here all the constant supermatrices: $\Lambda$\,, $\tau_3$\,, $k$\,,
${\rm K}$ and ${\rm C}$\,, as well as the supermatrix fields $Q$ and
$T$ are $8\times 8$ matrices defined on the retarded/advanced,
bosonic/fermonic and time-reversal sector (for the details of which
we refer to Ref.~\cite{Efetov97}). In Eq.~(\ref{propagator1}) ${\hat
{\cal L}}_{\rm a} =
{\bf p}\cdot (\partial_{\bf r}+[i{\bf
a}\tau_3,\,])-\partial_{\bf r}V\cdot
\partial_{\bf p}$ is the covariant Liouvillian,
where the supermatrix ${\bf a}={\rm diag}({\bf k}_+,{\bf k}_-)^{\rm
ra}$ with the superscript standing for the retarded/advanced sector.
Eqs.~(\ref{propagator1}) and (\ref{freeenergy}) are the exact
starting point of the succeeding analysis.

We remark that the coarse graining $\mathbb{T}$\,, with the element
commensurating with the uncertainty intrinsic to wave dynamics
\cite{Altland07}, suffices to regularize the field theory. (The
results below, however, are insensitive to the details which,
therefore, are not presented.) Indeed, the Liouvillian ${\hat {\cal
L}}_0$ induces a deterministic flow on $\mathbb{T}$\,: $X\mapsto
X(t)\equiv ({\bf r}(t),{\bf p}(t))$\,. The flow (i) has positive
$\lambda$
and is ergodic and, (ii) enjoys exponential decay of correlations
and the central limiting theorem for sufficiently regular
(H$\ddot{\rm o}$lder-continuous) functions. They lead to singular
hydrodynamic (steady) states \cite{Gaspard96} (and justify the
application of the BGS conjecture in the qualitative discussions),
as we also show below. Then, associated with the coarse graining
$\mathbb{T}$ the Gibb's entropy production results which can be
shown to be positive-definite \cite{Gaspard96}, signaling a stable
field theory.

Notice that properties (i) and (ii) follow directly from the choice
of the potential $V$ \cite{potential}. Indeed, the former is given
by the Donnay-Liverani theorem \cite{Donnay91}. To justify the
latter we introduce the Birkhoff coordinate: $(\theta,\varphi)$\,,
where $\theta$ is the position of the incident ray at the dielectric
cylinder boundary ($r=r_0$), and $\varphi\in [0,\pi]$ is the angle
made between the incident ray and the tangential direction of the
cylinder boundary. Then, we observe that the rotation function
$\Delta \theta(\varphi)$ defined in \cite{rotationfunction}, as well
as its first-order derivative, is discontinuous at
$\varphi=\frac{\pi}{2}$ while continuous at each interval. From the
B$\acute{\rm a}$lint-T$\acute{\rm o}$th theorem \cite{Donnay91}
property (ii) immediately follows.

{\it Hydrodynamic relaxation and Gaspard nonequilibrium steady
state}---Let us start from the ray optics limit: $\Delta /\lambda
\rightarrow 0$ and present a field-theoretic proof of the {\it
deterministic} diffusion. The emergence of hydrodynamic relaxation
from the Liouvillian dynamics is a long standing problem
\cite{Prigogine62}. It has been exactly solved for the periodic
Lorentz gases \cite{Gaspard96} as discussed here. To calculate
Eq.~(\ref{propagator1}) we employ the so-called rational
parametrization: $T=1+iW$\,. Here $W\Lambda+\Lambda W=0$ and,
moreover, $W(X)=-{\rm K}{\rm C}W^{\rm T}({\bar X}){\rm C}^{\rm
T}{\rm K}$\,. Inserting it into the action $F$ the perturbative
expansion in $W$ shows that $W^2$ is order of $\Delta/\lambda$\,.
Because of $\Delta /\lambda \rightarrow 0$ in the expansion of both
$F$ and the prefactor only the quadratic terms are to be kept. Upon
fixing the $U(1)$ gauge the succeeding calculations are formally
parallel to disordered systems \cite{Efetov97}, but, there is a
crucial difference namely the resolvent: $\{-i\omega^+ + {\hat {\cal
L}}_0 + {\bf p}\cdot i{\bf g}\}^{-1}$\,. Recall that, in disordered
systems, this field propagator is replaced by the heat kernel
recovering the diffusion. Upon passing to the time domain the
resolvent, denoted as ${\cal D}_{{\bf g},t}$\,, solves
\begin{eqnarray}
{\cal D}_{{\bf g},t} = e^{-t{\hat {\cal L}}_0} - \int_0^t d\tau
e^{-(t-\tau){\hat {\cal L}}_0} {\bf p}\cdot i{\bf g}\, {\cal
D}_{{\bf g},\tau} \,. \label{propagator2}
\end{eqnarray}

For low-lying modes the functional equation (\ref{propagator2})
gives
\begin{eqnarray}
{\cal D}_{{\bf g},t} = \bigg\{1- \int_0^t d\tau {\bf p}(-\tau)\cdot
i{\bf g}  \qquad \qquad \qquad \qquad \qquad
\label{classicalnonequilibrium}\\
- \int_0^t \! d\tau \int_0^{t-\tau} \! d\tau' {\bf g}{\bf g}:{\bf
p}(-\tau){\bf p}(-\tau-\tau') \bigg\}\, e^{-t {\hat {\cal L}}_0}
\nonumber
\end{eqnarray}
with the second-order iteration. Combined with property (i)
Eq.~(\ref{classicalnonequilibrium}) shows that, given a generic
initial distribution, for vanishing ${\bf g}$ it converges to the
invariant microcanonical measure in the limit $t\rightarrow
+\infty$\,. A small concentration gradient $i{\bf g}$ drives the
system out of equilibrium:
\begin{eqnarray}
{\cal D}_{{\bf g},t\rightarrow +\infty} = 1- \int_0^{+\infty} d\tau
{\bf p}(-\tau)\cdot i{\bf g}  + o({\bf g}^2)\,, \label{steadystate}
\end{eqnarray}
where the second term is the singular microscopic nonequilibrium
steady state found by Gaspard \cite{Gaspard96}. Furthermore, by
inserting Eq.~(\ref{classicalnonequilibrium}) into
Eq.~(\ref{propagator1}) and returning back to the frequency domain,
we find ${\cal Y}({\bf q},\omega) = 2\pi\nu/(-i\omega+D_{cl} {\bf
q}^2)$ and, with the help of property (ii), rigorously justify the
Green-Kubo formula:
\begin{eqnarray}
D_{cl} = \frac{1}{2} \int_{\mathbb T} dX \int_0^{+\infty} dt\, {\bf
p}(t)\cdot {\bf p}(0)\,.
\label{GreenKubo}
\end{eqnarray}
Eqs.~(\ref{steadystate}) and (\ref{GreenKubo}) confirm that the
normal diffusion appears as the Pollicott-Ruelle resonance as shown
earlier \cite{Gaspard96}. There, a completely different
technique--the $\zeta$-function theory--is used informing its
far-reaching relation to the field-theoretic approach. In obtaining
the diffusive two-point correlator the integration over $X$ and $X'$
is essential by which the scaling regime is recognized.

{\it Weak localization}---For $\Delta /\lambda \ll 1$ wave
interference accounts for fluctuations. Keeping the $W$-expansion up
to the fourth order the resolvent is modified to be $\{-i\omega^+ +
{\hat {\cal L}}_0 +\delta {\hat {\cal L}} +{\bf p}\cdot i{\bf
g}\}^{-1}$\,. Here the perturbed operator $\delta {\hat {\cal L}}$
is explicitly expressed as
\begin{eqnarray}
\delta {\hat {\cal L}}_{XX'} &=& - \frac{1}{\cal N}\sum_{{\bf g}\in
\mathbb{B}.\mathbb{Z}.}\int_0^{+\infty} \!\! dt e^{-i\omega^+t}
\langle
{\bar X} | {\cal D}_{{\bf g},t} |X\rangle \nonumber\\
&& \times \{\langle X|X'\rangle - \delta (X-X')\}
\label{decomposition}
\end{eqnarray}
which, in the ray optics limit, is nullified by the fine-grained
phase space structure because the bases $\langle X|$ and
$|X'\rangle$ satisfy $\langle X|X'\rangle = \delta (X-X')$\,. In
wave dynamics such point-like structure is smeared by the
uncertainty. Indeed, a famous theorem of periodic Lorentz gases
suggests that the bases $|X\rangle$/$\langle X|$ must instead be
constructed as a distribution supported by a small neighborhood of
$X$ and normalized to unity, with $\langle X|X'\rangle$ defined as
the overlap between two coarse grained representing points $X$ and
$X'$ \cite{Sinai85}. (Notice that the point-like distribution cannot
be spanned by these bases.) Furthermore, it was rigorously shown
\cite{Sinai85} that $|X\rangle$ relaxes to the microcanonical
measure at some short time scale ($\gtrsim \lambda^{-1}$) when the
support of $|X\rangle$ uniformly covers ${\mathbb T}$\,. Upon
combined with Eq.~(\ref{classicalnonequilibrium}) the matrix element
$\langle {\bar X} | {\cal D}_{{\bf g},t}|X\rangle$\,, in turn,
acquires a universal value while vanishes at shorter times
$t\lesssim \lambda^{-1}$. (For chaotic cavities this fact was
noticed by M{\"u}ller and Altland \cite{Altland07}.) Importantly,
such an interference correction conserves the probability (obeying
the so-called Ward identity) guaranteed by $\int_{\mathbb T} dX
\{\langle X|X'\rangle - \delta (X-X')\} = \int_{\mathbb T} dX'
\{\langle X|X'\rangle - \delta (X-X')\} =0$\,.

Repeat the same procedure of
Eqs.~(\ref{propagator2})-(\ref{GreenKubo}) taking into account the
third order iteration. After tedious calculations we arrive at a
similar diffusive two-point correlator but, crucially, $D_{cl}$ is
renormalized into
\begin{eqnarray}
 D_0(\omega) &=& D_{cl}\left[1-{1\over {\pi \nu}}\,
 \int\!\! \frac{d {\bf q}} {(2\pi)^2}\,
 \frac{1}{-i\omega+D_{cl} {\bf q}^2}\right]
\label{D}
\end{eqnarray}
for $\Delta \lesssim \omega \ll \lambda$\,. The second term fully
agrees with earlier theoretical prediction for periodic Lorentz
gases \cite{Tian05}, and resembles the well-known weak localization
correction of disordered systems. In the latter systems strong
localization develops for lower $\omega$\,. Nevertheless here such a
scenario breaks down which we now come to study.

{\it Diffusive-ballistic transport crossover}---For $\omega\ll
\lambda$ the zero mode (uniform component) of the $T$ field, denoted
as $T_0$ is established recognizing the microcanonical measure over
$\mathbb{T}$\,. Therefore, we factorize $T$ as $T=T_0 T_>$ and
insert it into the action, where $T_>$ are the fluctuating
component. With $T_>$ integrated out Eq.~(\ref{propagator1}) gives
\begin{eqnarray}
{\cal Y}({\bf q}\rightarrow 0,\omega) = \frac{(\pi\nu)^2}{64 {\cal
N}} \tilde{\sum_{{\bf
k}_\pm}} \int DQ_0\, e^{-{\tilde F}[Q_0,{\bf a}]} \qquad \qquad \nonumber\\
\times {\rm str}
\left[k(1+\Lambda)(1-\tau_3)Q_0k(1-\Lambda)(1-\tau_3)Q_0\right]\,,
\label{propagatorfield}
\end{eqnarray}
where $Q_0=T_0\Lambda T_0^{-1}$ and the zero mode action: ${\tilde
F}[Q_0,{\bf a}]=\frac{\pi\nu}{8}\, {\rm str}\{D^* [i{\bf
a}\tau_3,Q_0]^2 + \frac{i\omega^+}{2} \Lambda Q_0\}$\,. Here
$D^*={\rm Re}\, D_0 (\max\{\omega,\Delta\})$ with
$\max\{\omega,\Delta\}$ playing the role of the infrared cutoff.

To proceed further we invoke the Efetov parametrization for $Q_0$
\cite{Efetov97}. Then, the hydrodynamic (i.e., ${\bf a}$-) expansion
for Eq.~(\ref{propagatorfield}) is performed and kept up to the
second order. The remaining integral can be exactly done giving
\begin{eqnarray}
{\cal Y}({\bf q}\rightarrow 0,\omega) =
\frac{2\pi\nu}{-i\omega+D(\omega) {\bf
q}^2} \,, \,\, \omega \ll \lambda\,, \nonumber\\
D(\omega)= D^*\left[1-\frac{\Delta^2}{2\pi \omega^2}(1-e^{2i\pi
\omega/\Delta})\right]. \label{propagatorresult}
\end{eqnarray}
Strikingly, the diffusive-ballistic transport crossover is universal
despite that the present unit cell drastically differs from a
disordered one \cite{Taniguchi93}. The fundamental structure
difference at the unit cell level profoundly affects the ray optics
and thereby the field theory construction. Notice that $D_{cl}$ is
now renormalized into $D^*$ which is not reported for periodic
disordered systems \cite{Taniguchi93}. These low-lying modes are
diffusive for sufficiently large frequencies, i.e., $\Delta \ll
\omega \lesssim \lambda$\,, while ballistic in the opposite limit,
i.e., $\omega\ll \Delta$ because of $D(\omega)\approx
D_0(\Delta)\Delta/(-i\pi\omega)$\,.

{\it Possible experimental observations}---Eqs.~(\ref{steadystate}),
(\ref{GreenKubo}), (\ref{D}) and (\ref{propagatorresult}) are the
main results of this work. To confirm the universal
diffusive-ballistic crossover might be within the reach of the
experimental scope of Ref.~\cite{Fishman07} utilizing the optical
induction technique \cite{Segev02}. The incident probe beam is of
$10\, \mu{\rm m}$ full-width at half-maximum. The refractive index
is required to vary over a scale of $100\, \mu {\rm m}$ which, we
expect, might be possible by tuning the optical interference
pattern. The output light intensity profile $I(x,y;z)$ then
determines the dispersion through $\sigma^2(z) =\int\!\!\!\!\int
dxdy\, (x^2+y^2) I(x,y;z)$\,. The latter is directly transferred
into $D(\omega)$ by
\begin{eqnarray}
\sigma^2(z)-\sigma^2(0) = \int \frac{d\omega}{\pi}\,
 \frac{1-e^{-i\omega
z}}{\omega^2}\,D(\omega) \,,
 \label{dispersion}
\end{eqnarray}
giving Eq.~(\ref{widthresult}) with Eq.~(\ref{propagatorresult})
inserted. Finally, we stress that the universality of high-frequency
transport relies on the normal diffusion in the ray optical limit.
Thus, the predictions are applicable for a large class of periodic
dielectric structures.
In particular, the weak localization is stronger for the
quasi-one-dimensional transverse plane.

We are grateful to M. Garst for a valuable contribution. We have
also enjoyed discussions with A. Altland, J. R. Dorfman, P. Gaspard,
F. Haake, and Z. Q. Zhang, and were encouraged by Bambi Hu. This
work is supported by Transregio SFB 12 of the Deutsche
Forschungsgemeinschaft, NNSF of China (No. 10334020 and 10574027),
and partly by the Hongkong Baptist University.


\begin{thebibliography}{}
\bibitem{potential} The refractive index inside the
cylinder is central symmetric and is denoted as $n(r)\,, r\in
(0,r_0)$\,. It satisfies a) $V\in C^3((0,r_0))$\,, b) $h'(r) > 0$
for all $r \in (0,r_0)$\,, c) $\lim_{r\rightarrow 0} r^2 V(r) =0,\,
V(r_0)=0\,, V'(r_0^-)<0$\,, d) $\Omega(r)< 0$ and e) $a-2r_0
> -2 / V'(r_0^-)$\,. Here $h(r)\equiv r^2
[k_\perp^2-2\Omega_0 n_0 \,V(r)]$ and $\Omega(r)\equiv
rh'(r)/h(r)$\,.
These conditions, together with the cylinder geometry are merely for
technical reasons. Furthermore, $a$ satisfies $2r_0<a\leq
\frac{4}{\sqrt 3}r_0$ so that the mean free path is bounded from
above.
\bibitem{rotationfunction} The rotation function is defined as
$ \Delta\theta(\varphi) \equiv 2r_0\cos \varphi \int_{{\hat
r}(\varphi)}^{r_0}\, r^{-1}\{h(r)-(r_0\cos \varphi)^2\}^{-1/2}dr
$\,,
where $\hat r(\varphi)$ is the minimal distance as the trajectory
approaches the cylinder center, and $h(r)$ is defined in
\cite{potential}.
\end{thebibliography}
\end{document}